\documentclass{ws-procs9x6}

\begin{document}

\title{The present and the future of cosmology with Gamma Ray Bursts}

\author{G. Ghirlanda \& G. Ghisellini}

\address{ Osservatorio Astronomico di Brera\\
 Via E. Bianchi 46\\ 
 I--23807 Merate (LC)\\ 
E-mail: ghirla@merate.mi.astro.it}

\maketitle

\abstracts{Gamma Ray Bursts are among the most powerful astrophysical
sources and they release up to $10^{54}$ erg, if isotropic, in less
than few hundred seconds. Their detection in the hard X/$\gamma$ ray
band (at energies $\ge$10 keV) and out to very high redshift
(z$\simeq$6.3) makes them a powerful new cosmological tool (a) to
study the reionization epoch, (b) to unveil the properties of the IGM,
(c) to study the present universe geometry and (d) to investigate the
nature and cosmic evolution of the dark energy. While GRBs will surely
help to understand the first two issues in the future, the present
link between GRBs and cosmology has been made {\it concrete} by the
recent discovery of a tight correlation between their rest frame
prompt and afterglow emission properties which allowed their use as
{\it standard candles}. }

\section{Introduction}

Since their discovery in the latest sixties, the observational picture
of GRBs have been enriched with new pieces of evidence relative to
their emission properties, distance scale and intrinsic
properties. Also the theoretical models which pretend to interpret
these results were refined (e.g. see \cite{zhang} for a recent
review).

The nature of GRB progenitors, the huge amount of energy emitted
during their prompt and afterglow phase, their relativistic nature and
their cosmological distance scale makes them one of the most {\em
interdisciplinary} field of research in modern astrophysics. Their
study, in fact, requires concepts and methods drawn from the field of
stellar evolution, of the physics of compact objects and of
relativistic plasmas, of radiative process and they also represent a
laboratory to test general relativity and a new tool for cosmology.

The last point gathered the interest of the scientific community since
the settlement of the cosmological distance scale of the class of long
duration (i.e. $\geq$2 sec) GRBs in 1997 \cite{dj}.  In fact, the
large luminosity of GRBs and their redshift extension ($z\sim6.3$ up
to Sept. 2005 \cite{kw}) makes them the brightest distant objects we
know of, rising our hopes to test  different cosmological models
and to study the nature of dark energy (\cite{gh1}),
i.e. some of the hottest topics of modern cosmology.

\section{From the isotropic energy to the collimation corrected energy}

Apparently GRBs are all but standard candles: for $\sim$ 30 bursts
with measured redshift $z$ and fluence $F$ (integrated over the rest
frame 1keV--10MeV energy range) the {\it isotropic equivalent energy}
results,
\begin{equation}
E_{\rm iso}=\frac{4 \pi d_{L}(z)^2}{(1+z)}F
\end{equation}
with an average value $\sim 10^{53}$ erg and a wide spread over
$\sim$3 orders of magnitudes.
\begin{figure}[ht]
\centerline{\epsfxsize=4.1in\epsfbox{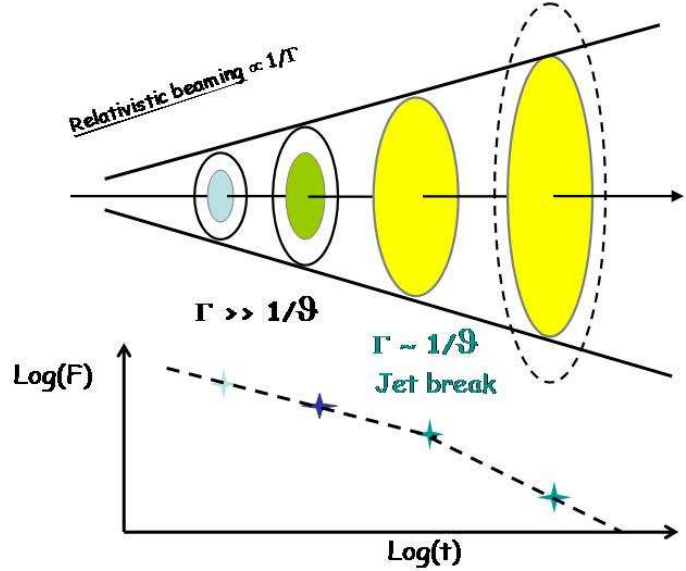}}   
\caption{Cartoon of the relation between the jet break time appearing
in the afterglow light curve (bottom panel) and the jet opening
angle. The shaded surfaces represent the area corresponding to the
beaming angle of the photons from which the observer perceives the
emitted radiation. When $1/\Gamma\sim\theta_{\rm jet}$ a change in the
decay slope of the afterglow light curve appears.\label{tbreak}}
\end{figure}

However, we know that GRBs are collimated sources with a typical
opening angle $\theta_{\rm jet}$ of few degrees \cite{frail}. The
estimate \cite{sari} of the jet opening angle is allowed by the
measure of the achromatic jet break time $t_{\rm break}$, i.e. the
time (typically between 0.5 and 6 days \cite{nava}) when the afterglow
light curve steepens. The link between $t_{\rm break}$ and
$\theta_{\rm jet}$ is shown in Fig.\ref{tbreak}. Due to the
relativistic beaming of the photons emitted by the fireball, the
observer perceives the photons within a cone with aperture
$\theta_{\Gamma}\propto 1/\Gamma$ where $\Gamma$ is the bulk Lorentz
factor of the material responsible for the emission. During the
afterglow phase the fireball is decelerated by the ISM and its bulk
Lorentz factor decreases, i.e. the beaming angle $\theta_{\Gamma}$
increases, with time. A critical time is reached when the beaming
angle equals the jet opening angle, i.e. $\theta_{\Gamma}\sim
1/\Gamma\sim\theta_{\rm jet}$, when the entire jet surface is visible
any further increase of the beaming angle (due to a decrease of the
bulk Lorentz factor) does not increase the visible surface any longer.
This time corresponds to a change (steepening) of the afterglow light
curve as schematically represented in Fig.\ref{tbreak}.

The measure of $\theta_{\rm jet}$ allows to correct $E_{\rm iso}$ and
derive the {\it collimation corrected energy},
\begin{equation}
E_{\gamma}=E_{\rm iso}(1-\cos\theta_{\rm jet})
\end{equation}
which is however spread over 2 orders of magnitudes and does not make
GRBs standard candles.

\begin{figure}[ht]
\centerline{\epsfxsize=5.0in\epsfbox{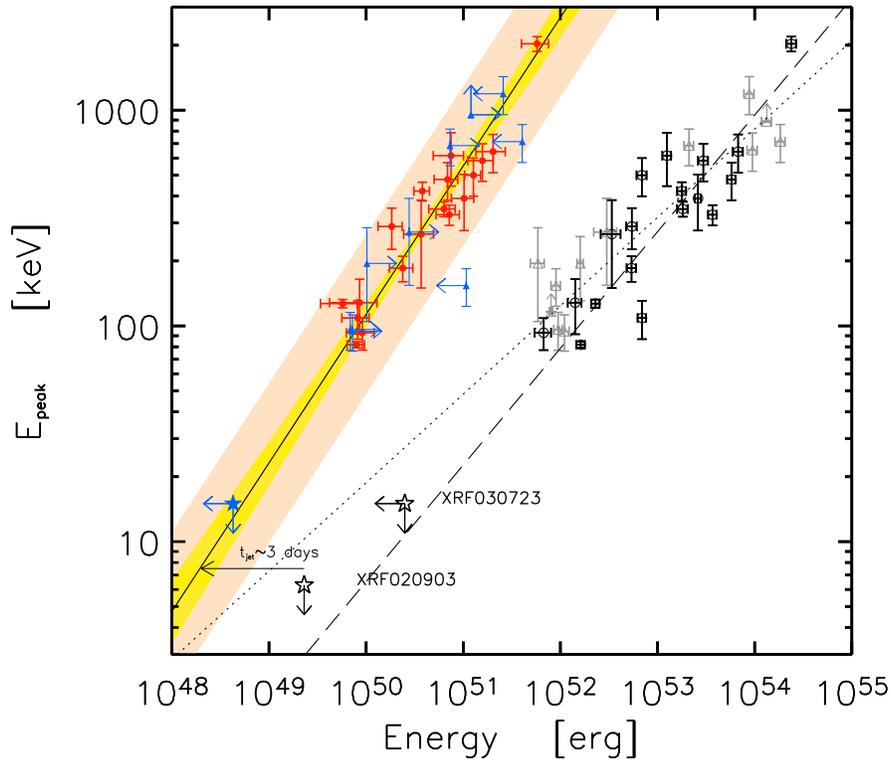}}   
\caption{Rest frame peak energy $E_{\rm peak}$ versus isotropic (open
symbols) and collimation corrected (filled symbols) energy.  The black
open circles are the 19 GRBs with measured $z$, $E_{\rm peak}$ and
$t_{\rm jet}$ for which the collimation corrected energy could be
computed (red filled circles).  Upper/lower limits on either one of
the variables are shown by the blue filled triangles.  The best fit
powerlaw to the red filled circles - Eq.(4) - is represented by the
solid line and its uncertainty by the thin yellow shaded region.  The
large (light orange) shaded region represents the $3\sigma$ gaussian
scatter of the data points around the correlation.  Also shown is the
$E_{\rm peak}-E_{\rm iso}$ correlation obtained by fitting the open
black data points and the open grey triangles (which are not
upper/lower limits) either accounting for the errors on both the
coordinates (long dashed line - slope = 0.54 and $\chi^2_{\rm r}=4.9$
for 27 dof) and by a linear regression (dotted line - slope = 0.4).
\label{ghirla} }
\end{figure}

The GRBs whose prompt emission spectrum is known over a wide energy
range so that their peak spectral energy $E_{\rm peak}$ (i.e. the peak
of the $EF_{E}$ spectrum) is properly constrained, present a
correlation \cite{amati} between $E_{\rm peak}$ and $E_{\rm
iso}$. With the present sample of 29 GRBs \cite{nava} (updated with
GRB~051022 - \cite{olive}) with accurately measured $z$, $E_{\rm
peak}$ and fluence we find (assuming a standard cosmology with
$\Omega_{\rm M}=0.3$, $\Omega_\Lambda=h=0.7$):
\begin{equation}
\left(\frac{E^\prime_{\rm p}} {100\, {\rm keV}}\right) \, =\,
(2.34\pm0.01)
\left(\frac{E_{\rm iso}} {2.76 \times 10^{52}\, {\rm erg}}\right)^{0.54\pm0.02}
\label{ama_new}
\end{equation}
with a reduced $\chi^{2}=4.97$ (27 dof) (Fig.\ref{ghirla}). The
scatter defined by the 29 data points in the $E_{\rm peak}$-$E_{\rm
iso}$ plane around the correlation defined in Eq.(\ref{ama_new}) is well
represented by a gaussian distribution with $\sigma=0.25$ dex.

If we instead correct the isotropic energy for the collimation angle
$\theta_{\rm jet}$ as defined in Eq.(2), a {\it tighter correlation}
between $E_{\rm peak}$ and $E_{\rm iso}$ is found \cite{gh}.

The GRBs for which the jet break time has been measured from the
afterglow light curve allows to derive the jet opening angles. To the
original sample of 15 bursts of \cite{gh} other 4 bursts have been
added: GRB~021004, GRB~041006, GRB~050525 and GRB~051022. The
correlation defined with this updated sample of 19 GRBs is:
\begin{equation}
\left(\frac{E^\prime_{\rm p}}  {100\, {\rm keV}}\right) \, =\,
(2.82\pm0.02)
\left(\frac{E_\gamma} {3.72 \times 10^{50}\, {\rm erg}}\right)^{0.69\pm0.04}
\label{ggl_new}
\end{equation}
which gives a better fit (with respect to Eq.(\ref{ama_new})), i.e. a
reduced $\chi^{2}=1.32$ (17 dof) (Fig.\ref{ghirla}). Also the scatter
defined by the 19 data points in the $E_{\rm peak}$-$E_{\rm iso}$
plane around the correlation defined in Eq.(\ref{ggl_new}) is well
represented by a gaussian distribution with $\sigma=0.1$ dex.

\section{Constraints on the cosmological parameters}

The tight $E_{\rm peak}$-$E_{\gamma}$ correlation makes GRBs standard
candles in the sense that, similarly to the stretching--luminosity
correlation of SNIa, allows to derive the GRB true energy. In fact,
the correlation represents the concrete possibility to use GRBs to
test the cosmological parameters and to study the nature of dark
energy.  In fig.3 we show the constraints on the
$\Omega_{M}$,$\Omega_{\Lambda}$ parameters obtained with the sample of
19 GRBs.

\begin{figure}[ht]
\centerline{\epsfxsize=4.5in\epsfbox{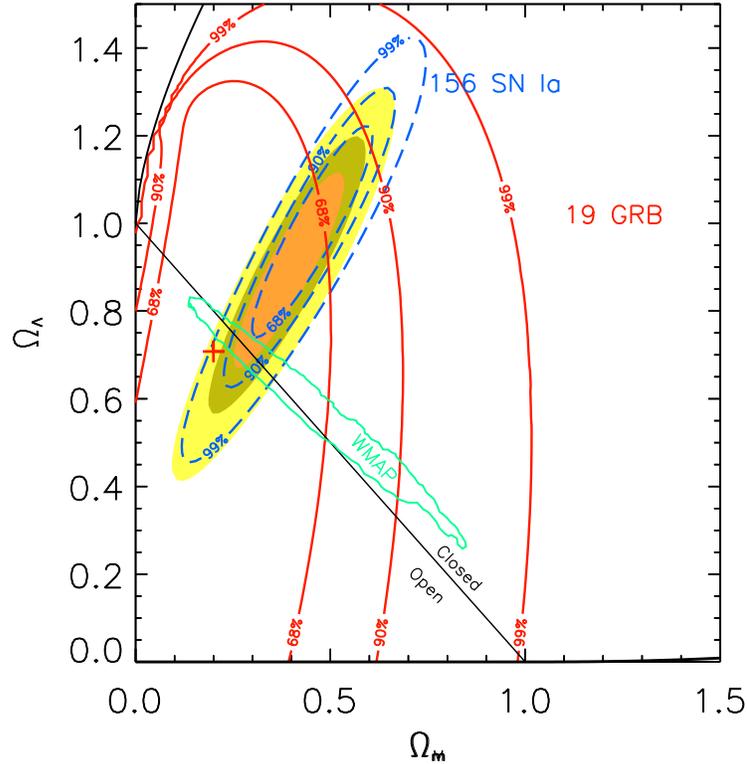}}  
\caption{Constraints (1,2 and 3$\sigma$) on the cosmological
parameters $\Omega_{M}$,$\Omega_{\Lambda}$ obtained with the 19 GRBs
with $z$, $E_{\rm peak}$ and $t_{\rm break}$ measured (solid
contours).  The contours obtained with the 156 SNIa of the Gold sample
are also shown (long--dashed contours) and the
combined GRB+SNIa contours (shaded regions) are represented. Also
shown are the 90\% constraints obtained with the WMAP data.
\label{ghi} }
\end{figure}

Although the contours obtained with 19 GRBs are shallow due to the low
number of bursts (only 19) compared to SNIa, we can see that GRB
constraints are complementary to those defined by SNIa and CMB in the
$\Omega_{M}$-$\Omega_{\Lambda}$ parameter space. GRBs are detected out
to very high redshifts (the present highest detected $z\sim6.3$ which
unfortunately has an unconstrained peak energy) and therefore they can
give better constraints in combination with the other cosmological
probes.

\section{The future of cosmology with GRBs}

Two urgent open issues related to the use of GRBs for cosmology are
(i) the circularity problem \cite{ghis} and (ii) the small number of
the present sample of GRBs with measured $z$, $E_{\rm peak}$ and
$t_{\rm break}$.

(i) The luminosity distance which allows to compute $E_{\rm iso}$ in
Eq.(1) (and $E_{\gamma}$ from Eq.(2)) is derived by assuming a
cosmological model (defined by the
$\Omega=(\Omega_{M},\Omega_\Lambda)$ parameters). Therefore, the slope
and normalization of the correlation, i.e. $E_{\rm peak}=K
E_{\gamma}^{g}$, depend on the particular cosmological model adopted,
i.e. $K(\Omega)$ and $g(\Omega)$, and we cannot use the correlation
found in a particular cosmology
(e.g. $\Omega_M=0.3$,$\Omega_\Lambda=0.7$) to derive the constraints
on the cosmological parameters. A method \cite{gh} that circumvents
this ``circularity'' problem is based on fitting the correlation in
every cosmology and finding the cosmologies (i.e. the $\Omega$ values)
which minimize the $\chi^2$ of the fit. A more refined method based on
Bayes theorem has been proposed and applied \cite{fir}. However these
methods have the inconvenience that the resulting constraints are less
stringent than in the case that the slope $g$ and normalization $K$ of
the correlation could be fixed (i.e. cosmology independent). The
latter situation can be realized either if the correlation is
calibrated with a sample of low redshift GRBs, i.e. at redshifts
$<0.1$, or if its theoretical interpretation is found. In both cases
in fact, the slope and normalization would become independent of
$\Omega$. We stress that the requirement of having low redshift GRBs
is not so stringent: we estimate \cite{gh2} that a sample of 12 GRBs
with redshift $z\in(0.9,1.1)$ can be used to calibrate the correlation
with a precision better than 1\% (i.e. the same accuracy acheived with
12 bursts with $z\in(0.45,0.75)$). A valid alternative is to find a
solid theoretical interpretation of the correlation which would fix
the values of $K$ and $g$ independently from the cosmological
parameters. 

(ii) A simulation of a sample of 150 GRBs shows \cite{gh2} that the
1$\sigma$ constraints obtained with GRBs alone might improve of
roughly a factor 10 with respect to the present constraints obtained
with only 19 bursts (solid contours in Fig.\ref{ghi}). However, the
increase of the sample of GRBs used in cosmology represents an
issue\cite{lamb} because, for every GRB which is added to the samepl
of 19 we need to measure: (i) the redshift $z$; (ii) the jet break
time $t_{\rm break}$ from the X--ray/Optical light curve (this measure
requires the monitoring of the GRB light curve up to $\sim 10$ -
depending on the redshift of the source $z$) (iii) the peak energy of
the prompt emission $E_{\rm peak}$

In particular the measure of $E_{\rm peak}$ requires to have a
spectrum of the prompt emission of the burst over a wide energy
range. In fact, we know that GRB spectra are characterized by a
featureless continuum which is often represented \cite{band} by a
double smoothly--joined powerlaw model. Typically the $EF_E$ spectrum
shows a peak at $\approx$ 300 keV \cite{preece} and it has been shown
to be extended to very low values (i.e. $\approx$ 20 keV) by the class
of X--Ray Flashes \cite{saka}. Therefore a detector which can cover
the 10 keV--1MeV energy range is required \cite{lamb} to constrain
$E_{\rm peak}$.

Among the presently flying instruments catching GRBs, Swift has the
highest detection rate ($\sim$ 1 GRB every 3 days roughly) as it was
specifically designed to detect and monitor the burst on--the--fly to
pinpoint its position with a high accuracy \cite{gher} (to better than
$\sim 5$'' in the X--ray band). However, the limited energy range
(15--150 keV) of the high energy instrument on board Swift does not
allow to constrain the peak energy of most GRBs it detects. However,
the IPN satellites and Hete--II \cite{lamb} have better chances to
measure the $E_{\rm peak}$ parameter and, as it appeared from the
recent case of GRB~051022, their joint work with Swift might be the
key to collect some more GRBs to be added to the presently small
sample of GRBs used as standard candles.

\section*{Acknowledgments}
A. Celotti, C. Firmani, D. Lazzati and F. Tavecchio are thanked for
continuous stimulating discussions.  The Italian MIUR is thanked for
funding (cofin grant 2003020775\_002)

\end{document}